\documentclass{webofc}

\usepackage[varg]{txfonts}   
\usepackage{hyperref}
\usepackage{url}
\hypersetup{colorlinks=true,citecolor=blue,urlcolor=blue,linkcolor=blue}
\usepackage{caption}
\usepackage{subcaption}
\usepackage{booktabs} 
\RequirePackage{orcidlink}
\begin{document}
\title{mkFit for track fitting with the CMS Phase-2 detector}
%
%

\author{
        \firstname{Leonardo} \lastname{Giannini}\inst{1}\orcidlink{0000-0002-5621-7706}\fnsep\thanks{\email{leonardo.giannini@cern.ch}} \and
        \firstname{Emmanouil} \lastname{Vourliotis}\inst{1}\orcidlink{0000-0002-2270-0492}\fnsep\thanks{\email{emmanouil.vourliotis@cern.ch}}
         		on behalf of the CMS Collaboration\footnote{Copyright 2026 CERN for the benefit of the CMS Collaboration. Reproduction of this article or parts of it is allowed as specified in the CC-BY-4.0 license.}}

\institute{University of California, San Diego}

\abstract{
The mkFit algorithm provides an implementation of the Kalman filter-based track reconstruction algorithm that exploits both thread- and data-level parallelism. 
It has been adopted by the CMS Collaboration as the primary track building algorithm during the Run 3 of the LHC for both the online and offline track reconstruction sequences. Thanks to mkFit, an average speedup of a factor 3.5 is achieved in track building while retaining or improving physics performance. As a consequence of the speedup provided by mkFit in track building, track fitting has become a comparably significant component of the overall tracking time. Given the increased demands of the High-Luminosity Large Hadron Collider, further speedup can be achieved by extending mkFit to the track fitting task. We present preliminary results for track fitting in the CMS High-Level Trigger using mkFit, covering both the physics and the computational performance, based on realistic Phase-2 simulations.
}

\maketitle
\section{Introduction}\label{intro}

The CMS track reconstruction proceeds through three main steps: track seeding, pattern recognition (or track building), and track fitting~\cite{CMS:2014pgm}, with both building and fitting based on Kalman filter techniques~\cite{Fruhwirth:1987fm}. A final track selection is then applied to the reconstructed candidates. 
Track building is the most complex and computationally demanding of these steps, while track fitting uses the same Kalman filter techniques, applied inside-out and outside-in, to obtain the most precise estimate of the track helix parameters.
In the traditional, “legacy” tracking, these steps are repeated in multiple iterations, where tracks are reconstructed progressing from the simplest to the most complex track topologies. Hits belonging to high-purity tracks identified in a given iteration are not considered in subsequent iterations to reduce combinatorics. Online and offline tracking have different numbers of iterations, due to the different requirements at each reconstruction level. \\

\begin{figure}[h!]
\centering
\includegraphics[width=0.99\textwidth]{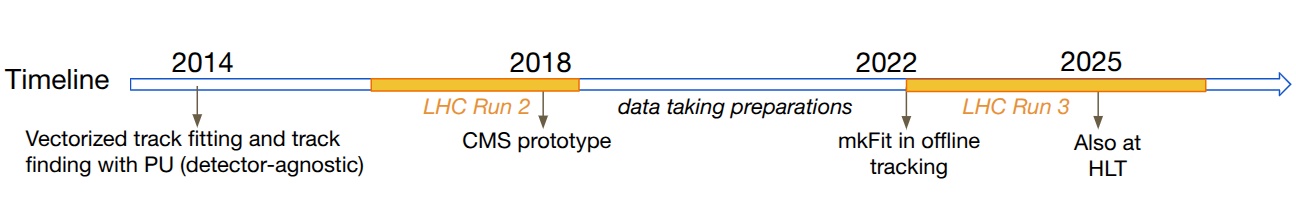}
\caption{Schematic timeline of mkFit developments.}
\label{timeline}
\end{figure}

Before the Run 3 of the LHC, the CMS track reconstruction relied on the Combinatorial Kalman Filter (CKF) algorithm~\cite{CMS:2014pgm} for both building and fitting. 
In Run 3, the mkFit algorithm~\cite{Lantz:2020yqe}, which exploits vectorization and parallelization on multi-core CPU architectures through Intel TBB and the matriplex library, was introduced for the track building step, first in offline reconstruction since 2022, and subsequently online, at the High Level Trigger (HLT), since 2025. The algorithm was initially developed for the vectorization of track fitting in a detector-agnostic framework, and was subsequently extended to track building. Further developments using the CMS detector geometry, led to the adoption by the experiment in Run 3, as shown by the schematic timeline in Fig.~\ref{timeline}.
Across the offline iterative tracking sequence, mkFit is used in the main iterations, accounting for approximately 90\% of all reconstructed tracks with $p_{\mathrm{T}} > 0.5$~GeV, with physics performance consistent with the legacy algorithms. 
The adoption of mkFit has reduced the average offline track building time by a factor of 3.5~\cite{CMS-DP-2022-018}, bringing it down to a level comparable to the track fitting time despite the greater combinatorial complexity of the building task. 
Summed over all iterations, this corresponds to an overall $1.7\times$ reduction in building time, a 25\% reduction in total tracking time, and a 10--15\% increase in event throughput, driven by the effective vectorization of about 70\% of operations and parallelization of over 95\% of the code, and by the general algorithm optimization.\\

The High-Luminosity LHC (HL-LHC) is expected to deliver an integrated luminosity of 3000~fb$^{-1}$ over ten years and an average pileup of about 200 collisions per event. Standard tracking algorithms scale super-linearly with pileup, motivating new solutions that can fully exploit the upcoming detector layout and the available computing resources. Substantial progress was recently shown in the scope of the CMS Phase-2 track reconstruction at HLT, through new algorithms such as Patatrack and LST, which change the legacy seeding and iterative tracking paradigms in the attempt to optimize the usage of both GPU and CPU architectures. The mkFit algorithm continues to be used for track building in this evolving picture~\cite{CMS-DP-2026-026}. In this context, extending mkFit to track fitting is a natural direction of development to further reduce the overall track reconstruction time. This note reports on the status of these developments~\cite{CMS-DP-2026-026,CMS-DP-2026-031}, using the new HLT tracking configuration as a testbed to evaluate mkFit as an alternative to the legacy CKF fitting algorithm.\\

\section{mkFit for track fitting in the CMS Phase-2 HLT track reconstruction}
\label{algo}

\begin{figure}[h!]
\centering
\includegraphics[width=0.55\textwidth]{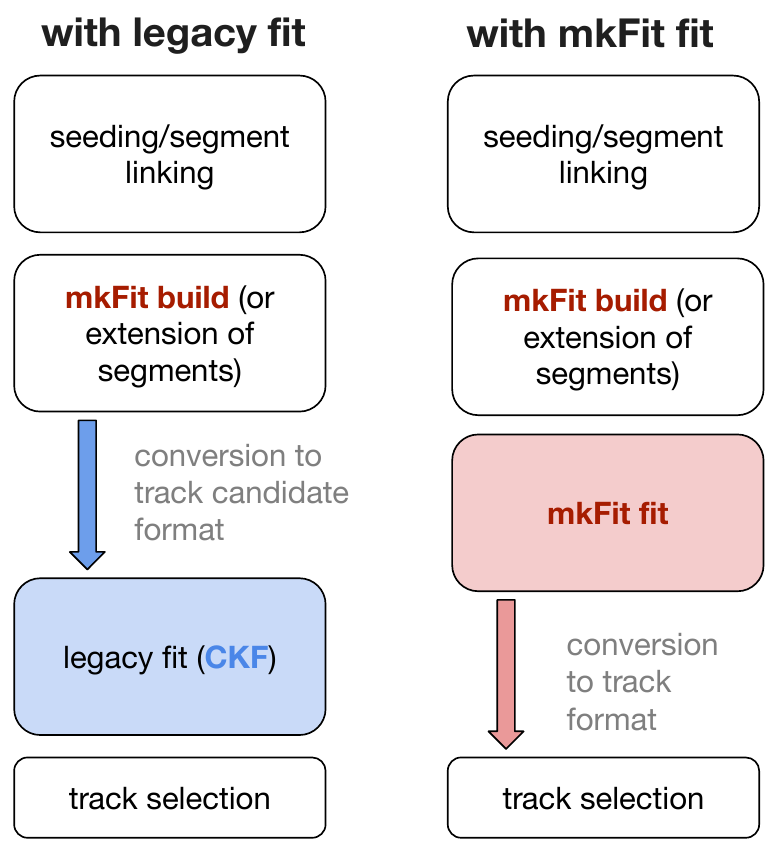}
\caption{Schematic representation of the tracking workflow when using the legacy fit (left) and the mkFit fit (right).}
\label{figs}
\end{figure}

The mkFit fitting algorithm code is designed to replace the legacy CKF fitting while retaining the same output data structures. 
The mkFit fit uses as input the mkFit built track candidates, with no need for data-format conversions that are instead required in the case of the legacy algorithm. After the fit, a conversion is performed, to return the same format as the legacy algorithm. Hence, the usage of mkFit for track fitting is transparent to downstream reconstruction sequences. A schematic representation of the track reconstruction tasks with the legacy and mkFit fits is shown in Fig.~\ref{figs}.\\

The fitting algorithm also follows closely the legacy CKF algorithm, but with a few simplifications.
Each track is fitted twice, first inside-out (forward) and then outside-in (backward), using as input the state and hit collection of the built track candidates. The hit outlier rejection is simplified compared to the legacy algorithm, and no “smoothing” is applied during the backward fit, whereas the legacy fit takes the weighted sum of forward and backward states corresponding to each hit in order to improve the precision of the estimate of the track parameters.\\

To exploit vectorization, tracks are grouped according to their hit multiplicity and fitted in parallel: for each group, the $n$-th hit is added simultaneously to all tracks in the group, with the matriplex library providing implicit vectorization of the underlying Kalman filter operations across a fixed number of tracks, $N_{\mathrm{V}}$, corresponding to the CPU operation vector width. Groups whose size is not a multiple of $N_{\mathrm{V}}$ are handled through a remainder group processed at the end. After the forward and backward fits are completed, hits are checked for compatibility with the track and outliers are identified through a $\chi^2$ selection. \\

The current, simplified outlier rejection criteria are tuned separately depending on the track $p_{\mathrm{T}}$. For tracks with $p_{\mathrm{T}} > 1$~GeV, a hit is rejected as an outlier if the sum of the given hit residual $\chi^2$ in the forward and in the backward fit prediction is larger than 20, and each $\chi^2$ (from the forward or backward fit prediction) is larger than 8. For tracks with $p_{\mathrm{T}} < 1$~GeV, the corresponding thresholds are 15 for the $\chi^2$ sum, and 7 for individual $\chi^2$ values. Tracks with at least one hit removed are regrouped according to their new hit multiplicity, and undergo a second fit without the rejected hits, yielding updated final track parameters. No further iteration of the hit outlier rejection is currently implemented.
This procedure allows mkFit to reproduce the fitting functionality of the legacy CKF fit, while retaining the performance benefits of a vectorized, parallel execution.

\section{Physics performance}
\label{phys}

The tracking performance is evaluated in terms of tracking efficiency, fake rate and resolution, using simulated samples where reconstructed tracks are associated with a simulated tracks. A reconstructed track is considered matched to a simulated track if more than 75\% of its hits originate from this simulated track. If this is not the case, the reconstructed track is marked as misidentified and referred to as a fake track. 

The tracking efficiency is defined as the fraction of simulated tracks associated with at least one reconstructed track, while the tracking fake rate is defined as the fraction of misidentified reconstructed tracks. The resolution is defined as the width of a Gaussian fit to the parameter response. Only simulated charged particles originating from the signal (hard scattering) vertex are used in the efficiency computation, with the selection applied to simulated tracks explicitly indicated in each figure; in addition to the simulated track transverse momentum ($p_{\mathrm{T}}$) and pseudorapidity ($\eta$), the radial and longitudinal displacement from the detector center, $r$ and $z$, are also used in these selections. For the fake rate and resolution computation, all simulated tracks are used regardless of their originating vertex.

The performances are compared across HLT tracking workflows where the same tracks are fitted either with CKF or with mkFit, using a simulated $t\bar{t}$ sample with 200 pileup interactions as the benchmark scenario, reconstructed with the current baseline tracking sequence proposed for the CMS HLT reconstruction in Phase-2~\cite{CMS-DP-2026-026}. The results shown here are obtained before the application of high-purity track selections.\\

\begin{figure}[h!]
\centering
    \begin{subfigure}[b]{0.45\textwidth}
    \centering
    \includegraphics[width=\textwidth]{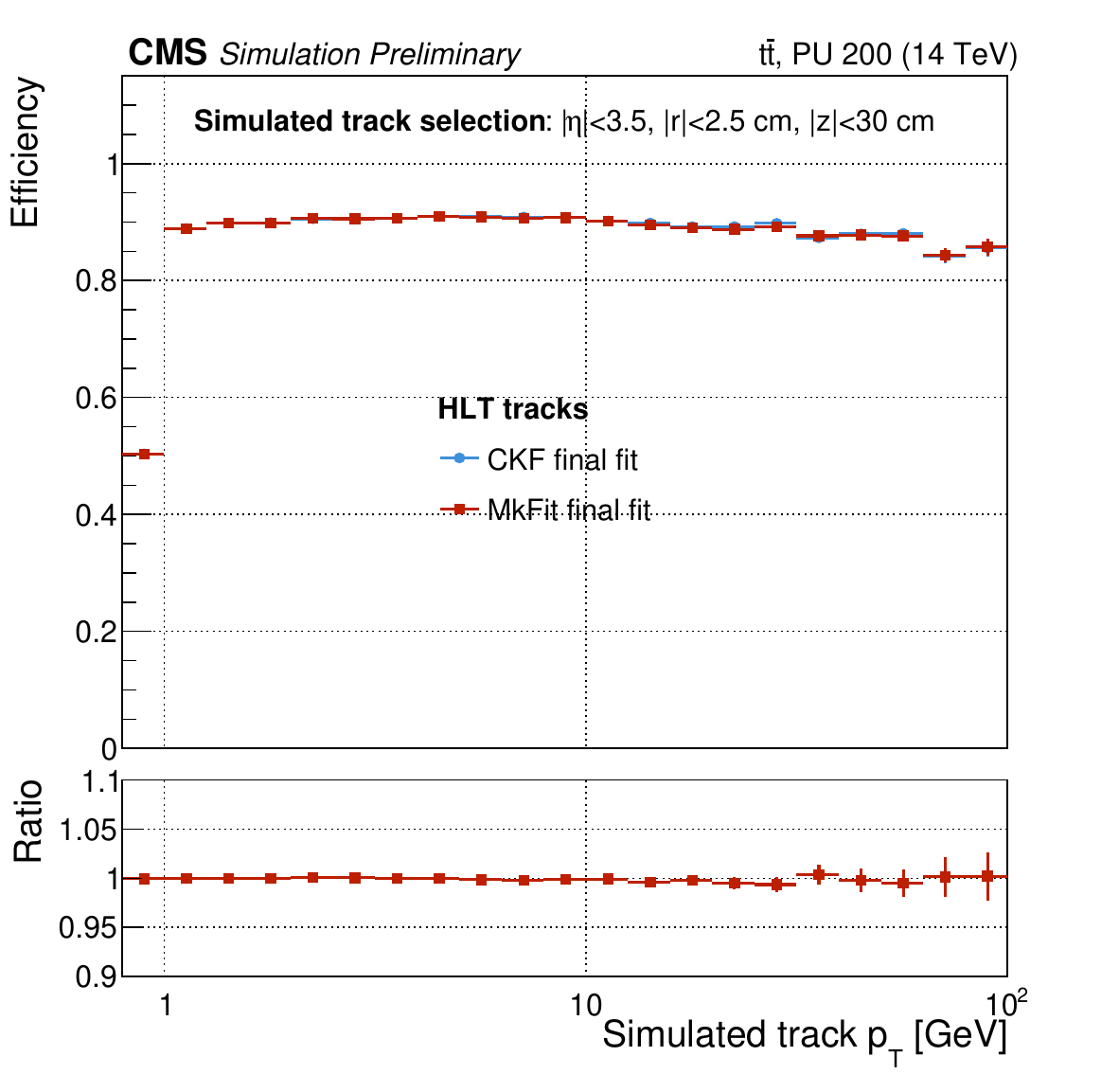}
    \caption{}\label{fig:image1}
    \end{subfigure}
\quad
    \begin{subfigure}[b]{0.45\textwidth}
    \centering
    \includegraphics[width=\textwidth]{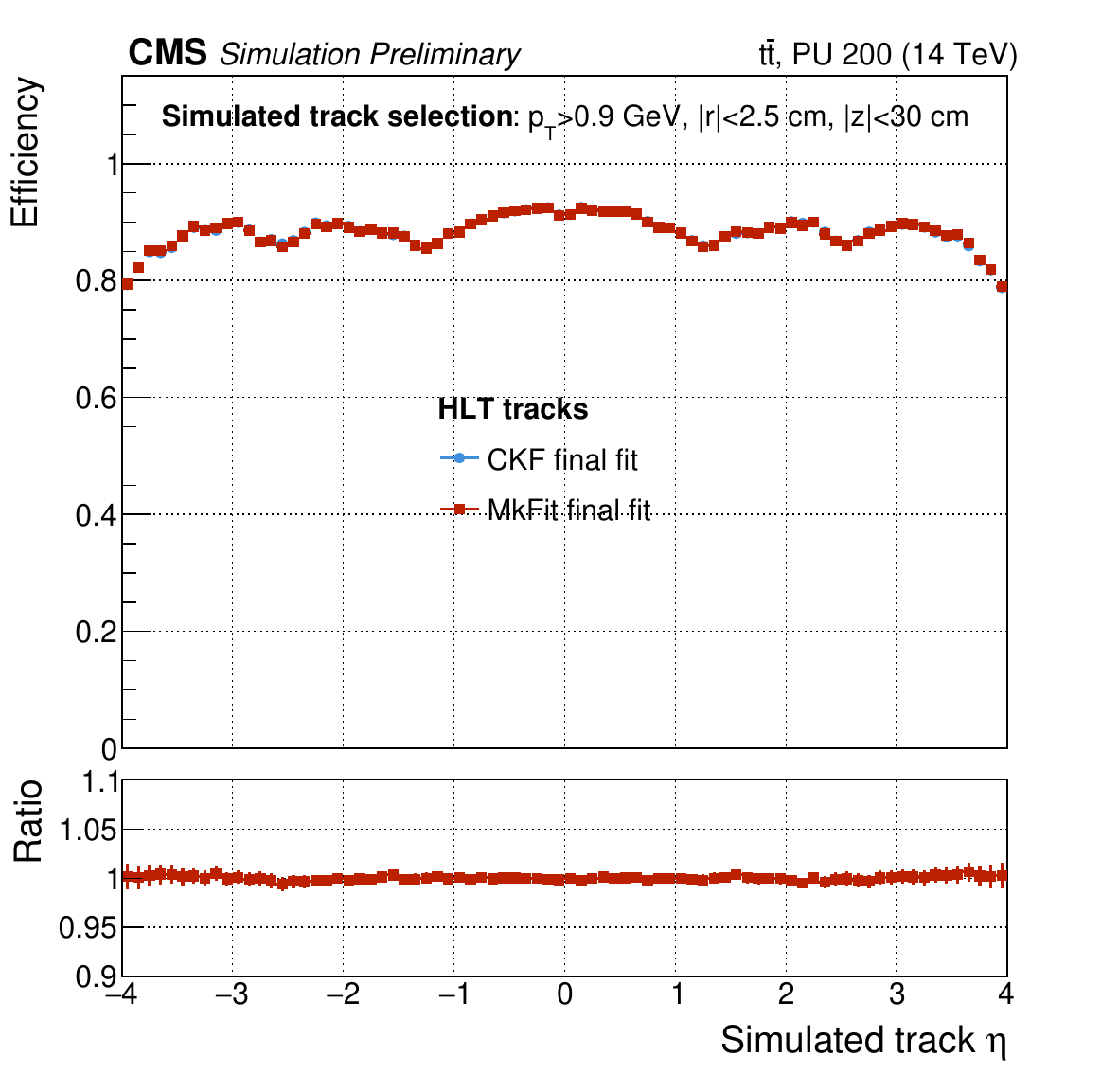}
    \caption{}\label{fig:image2}
    \end{subfigure}
\\
    \begin{subfigure}[b]{0.45\textwidth}
    \centering
    \includegraphics[width=\textwidth]{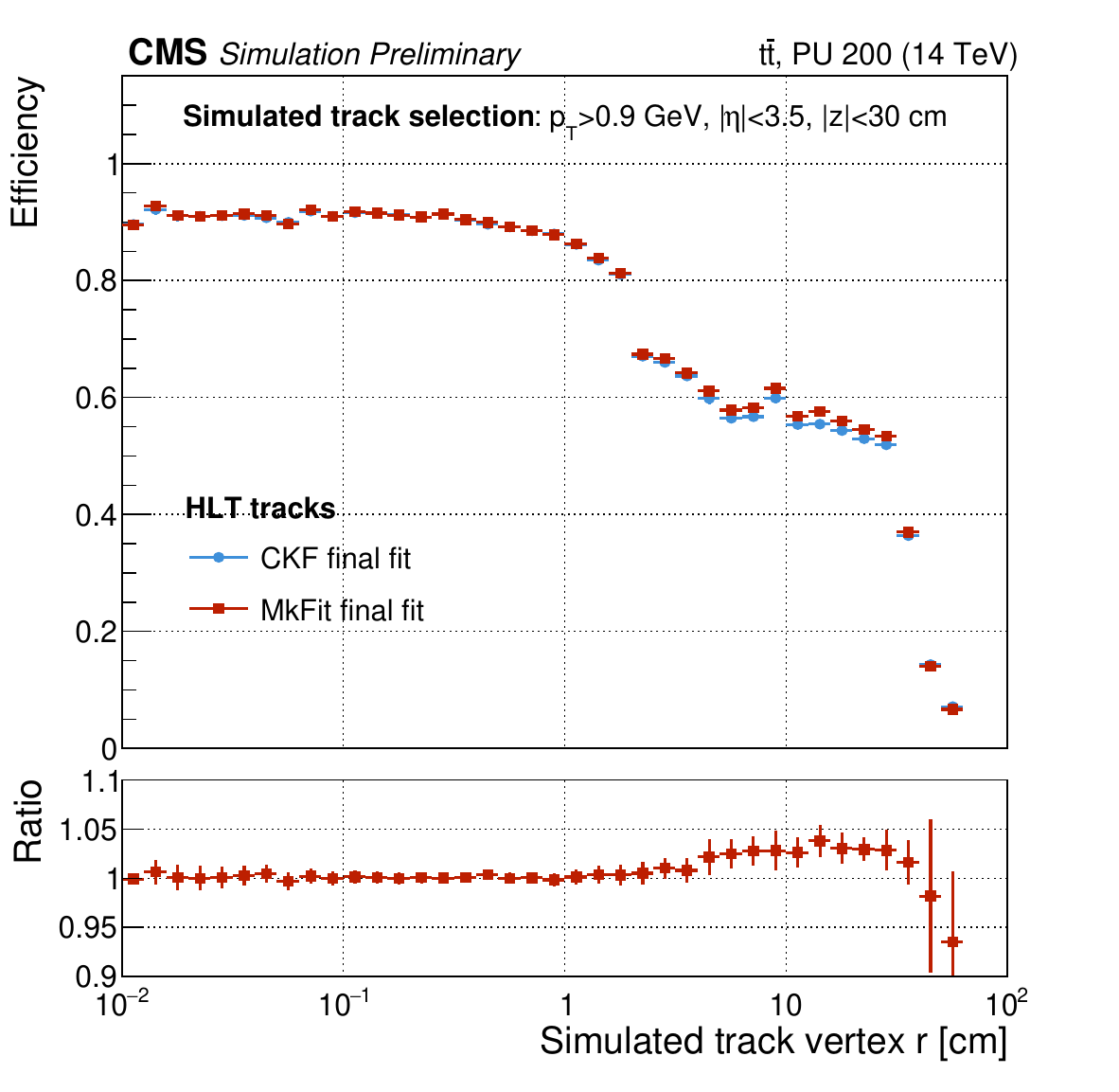}
    \caption{}\label{fig:image3}
    \end{subfigure}
\caption{Tracking efficiency for HLT reconstructed tracks after the CKF (blue) and mkFit (red) final fits, as a function of the simulated track $p_{\mathrm{T}}$ (\subref{fig:image1}), pseudorapidity $\eta$ (\subref{fig:image2}) and production radius $r_{\mathrm{vertex}}$ (\subref{fig:image3}). Selections applied to the simulated tracks are reported in each panel.}
\label{fig:eff}
\end{figure}

The tracking efficiency for HLT reconstructed tracks after the CKF (blue) and mkFit (red) final fits is shown in Fig.~\ref{fig:eff}, as a function of the simulated track $p_{\mathrm{T}}$ for tracks with $|\eta| < 3.5$, $|d_0| < 2.5$~cm and $|d_\mathrm{z}| < 30$~cm (\subref{fig:image1}); as a function of the simulated track pseudorapidity $\eta$ for tracks with $p_{\mathrm{T}} > 0.9$~GeV, $|d_0| < 2.5$~cm and $|d_\mathrm{z}| < 30$~cm (\subref{fig:image2}); and as a function of the production radius $r_{\mathrm{vertex}}$ for tracks with $p_{\mathrm{T}} > 0.9$~GeV, $|\eta| < 3.5$ and $|d_\mathrm{z}| < 30$~cm (\subref{fig:image3}). The tracking efficiency is comparable between the two fitting algorithms.\\

\begin{figure}[htpb!]
\centering
    \begin{subfigure}[b]{0.45\textwidth}
    \centering
    \includegraphics[width=\textwidth]{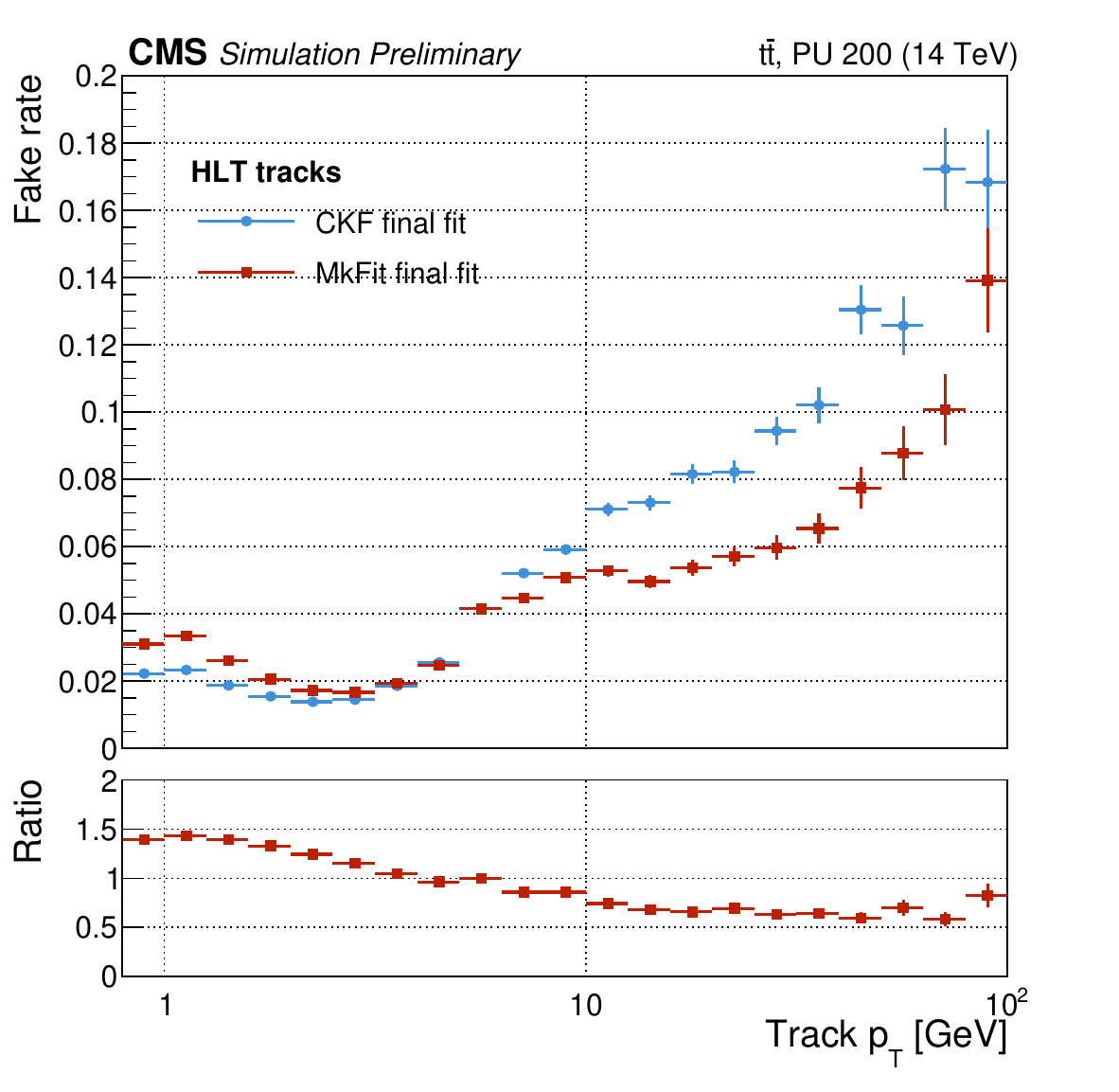}
    \caption{}\label{fig:fake_pt}
    \end{subfigure}
\quad
    \begin{subfigure}[b]{0.45\textwidth}
    \centering
    \includegraphics[width=\textwidth]{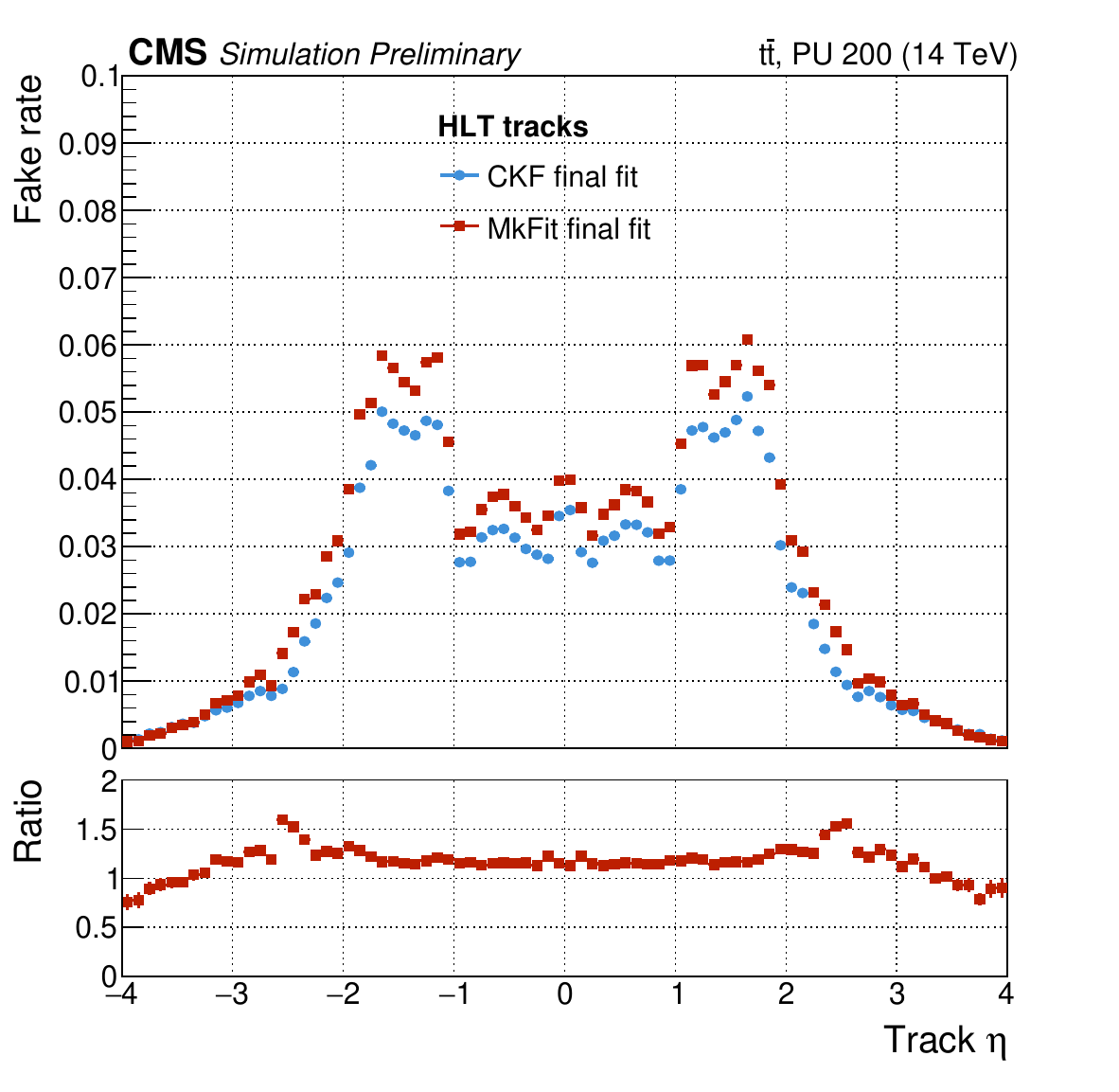}
    \caption{}\label{fig:fake_eta}
    \end{subfigure}
\caption{Tracking fake rate for HLT reconstructed tracks after the CKF (blue) and mkFit (red) final fits, as a function of the reconstructed track $p_{\mathrm{T}}$ (\subref{fig:fake_pt}) and $\eta$ (\subref{fig:fake_eta}).}
\label{fig:fake}
\end{figure}

The tracking fake rate for HLT reconstructed tracks after the CKF (blue) and mkFit (red) final fits is shown in Fig.~\ref{fig:fake}, as a function of the reconstructed track $p_{\mathrm{T}}$ (\subref{fig:fake_pt}) and as a function of the reconstructed track $\eta$. At low $p_{\mathrm{T}}$, the fake rate is higher for tracks fitted with mkFit by up to a factor of 1.5, although the fake rate remains below 4\% in this region; for $p_{\mathrm{T}} > 10$~GeV, the fake rate is instead reduced by almost a factor of 2 relative to CKF. As a function of $\eta$, the fake rate is slightly higher for tracks fitted with mkFit than for those fitted with CKF. These differences arise from the different tuning of the outlier rejection between the two fitting algorithms.\\

\begin{figure}[h!]
\centering
    \begin{subfigure}[b]{0.45\textwidth}
    \centering
    \includegraphics[width=\textwidth]{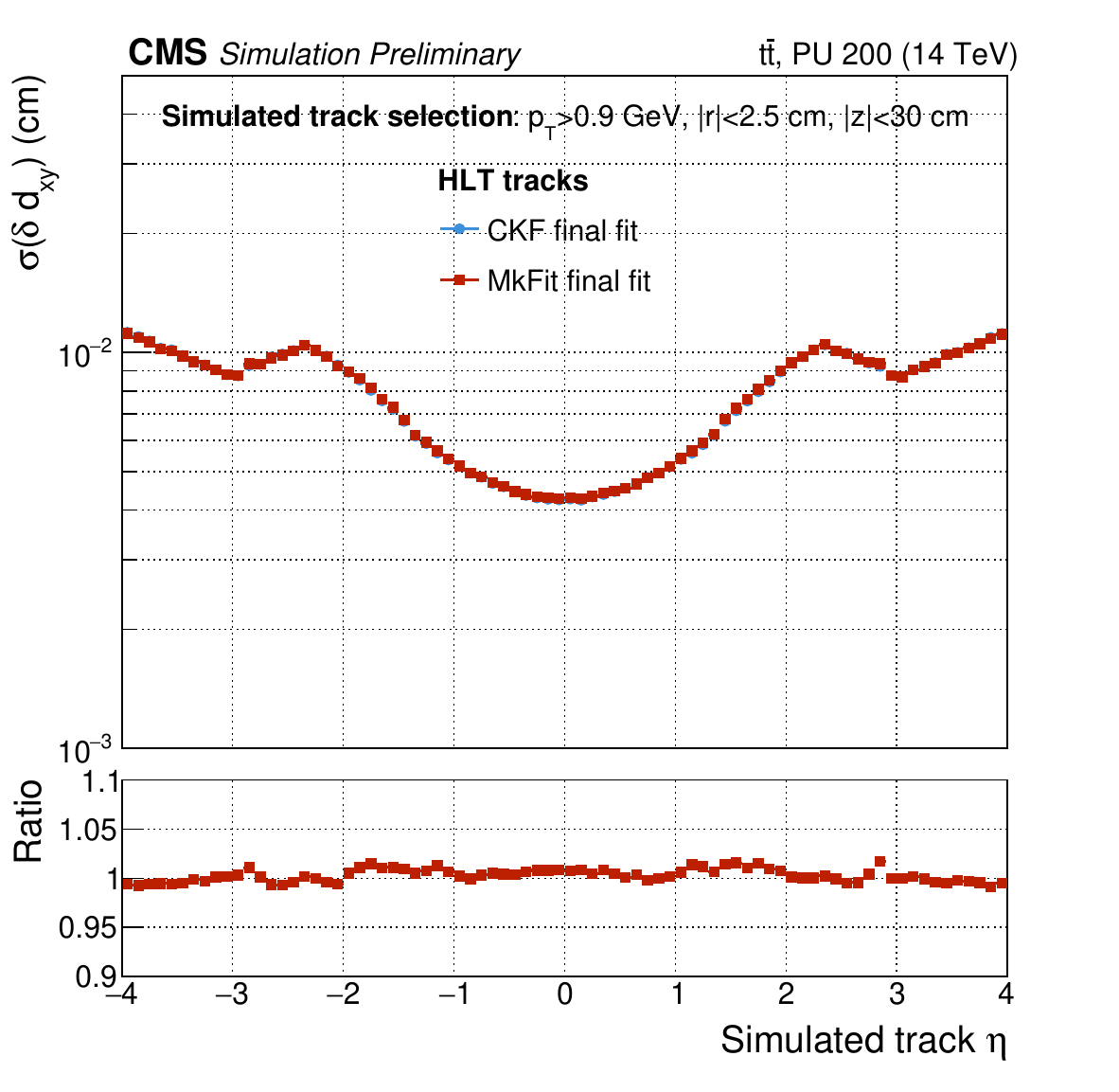}
    \caption{}\label{fig:res_dxy}
    \end{subfigure}
\quad
    \begin{subfigure}[b]{0.45\textwidth}
    \centering
    \includegraphics[width=\textwidth]{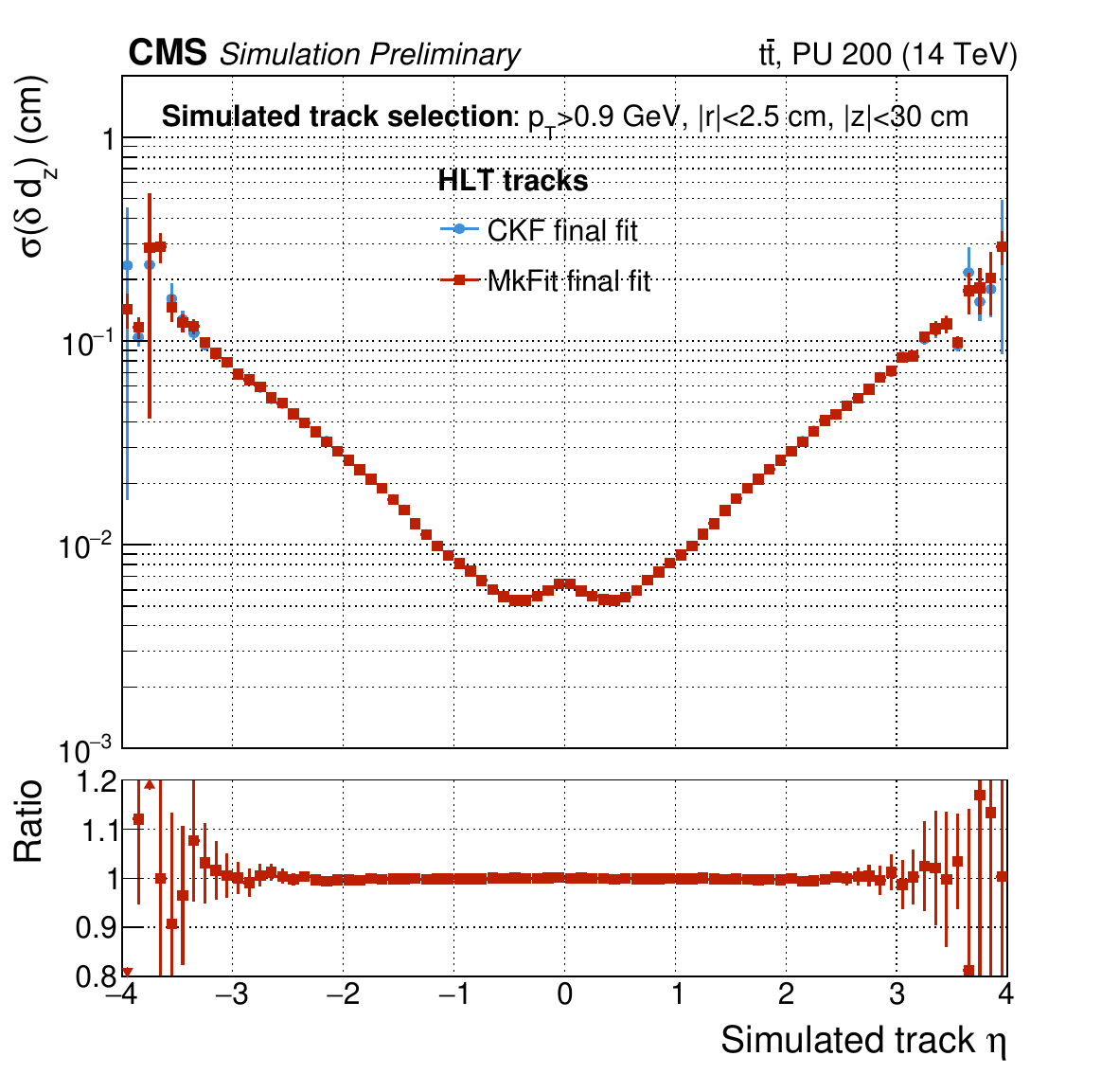}
    \caption{}\label{fig:res_dz}
    \end{subfigure}
\\
    \begin{subfigure}[b]{0.45\textwidth}
    \centering
    \includegraphics[width=\textwidth]{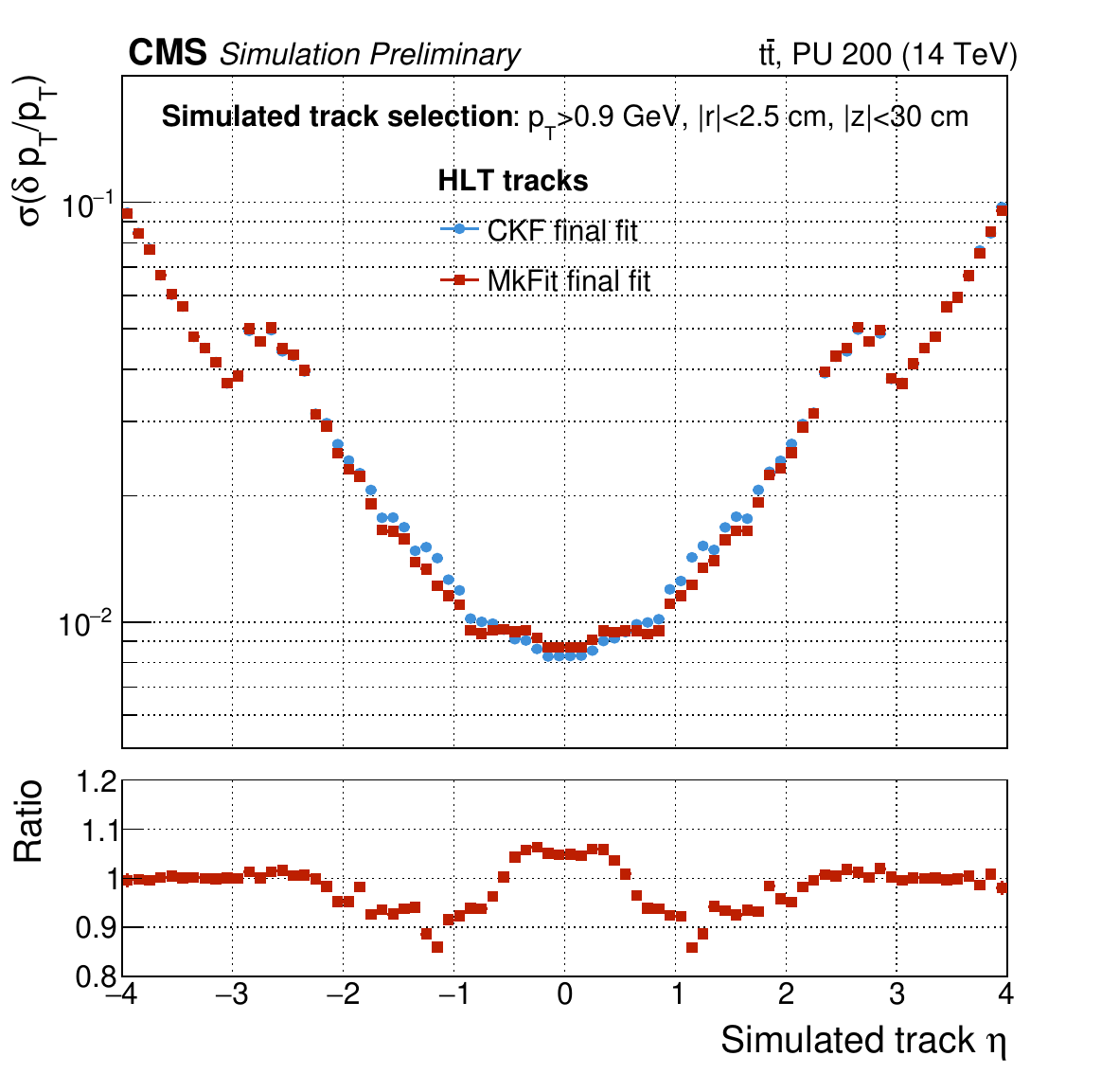}
    \caption{}\label{fig:res_pt_eta}
    \end{subfigure}
\quad
    \begin{subfigure}[b]{0.45\textwidth}
    \centering
    \includegraphics[width=\textwidth]{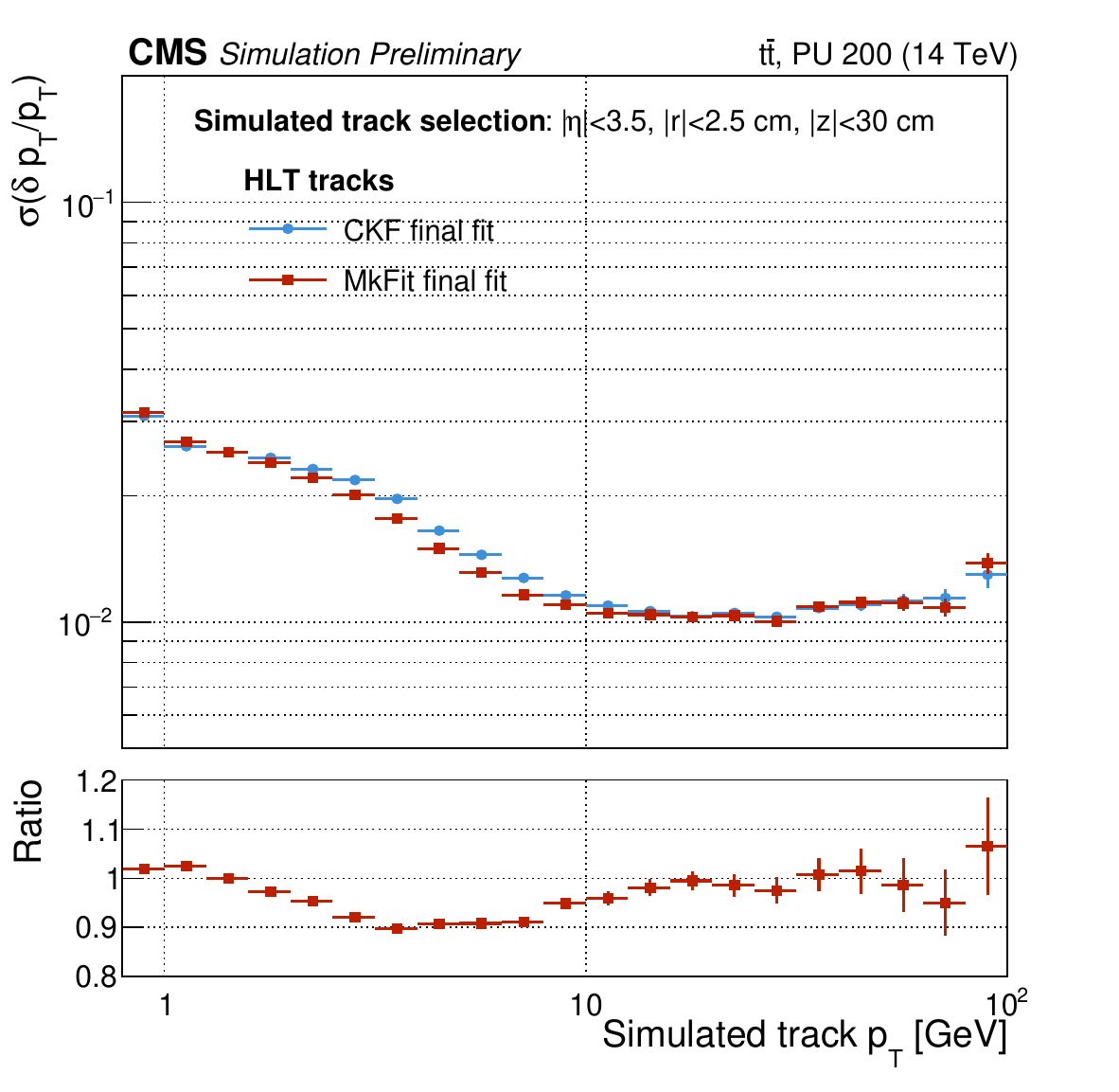}
    \caption{}\label{fig:res_pt_pt}
    \end{subfigure}
\caption{Track parameter resolutions for HLT reconstructed tracks after the CKF (blue) and mkFit (red) final fits: $d_\mathrm{xy}$ (\subref{fig:res_dxy}), $d_\mathrm{z}$ (\subref{fig:res_dz}) and $p_{\mathrm{T}}$ (\subref{fig:res_pt_eta}) resolution as a function of the simulated track $\eta$, and $p_{\mathrm{T}}$ resolution as a function of the simulated track $p_{\mathrm{T}}$ (\subref{fig:res_pt_pt}).}
\label{fig:resolution}
\end{figure}

The track parameter resolutions for HLT reconstructed tracks after the CKF (blue) and mkFit (red) final fits are shown in Fig.~\ref{fig:resolution}. The resolutions in $d_\mathrm{xy}$, $d_\mathrm{z}$ and $p_{\mathrm{T}}$ are shown as a function of the simulated track pseudorapidity $\eta$ (\subref{fig:res_dxy}, \subref{fig:res_dz}, \subref{fig:res_pt_eta}), and the $p_{\mathrm{T}}$ resolution is additionally shown as a function of the simulated track transverse momentum $p_{\mathrm{T}}$ (\subref{fig:res_pt_pt}). The resolutions in $d_\mathrm{xy}$ and $d_\mathrm{z}$ are consistent across the two algorithms in the full $\eta$ range, while the ratio of the $p_{\mathrm{T}}$ resolutions between mkFit and CKF shows a modulation as a function of $\eta$. As a function of $p_{\mathrm{T}}$, the $p_{\mathrm{T}}$ resolution ratio shows a slight variation but remains consistent within 10\% between the two algorithms. Further studies are ongoing to reach a complete physics assessment.\\

\section{Timing performance}
\label{time}

Timing measurements are performed to estimate the computing performance of the HLT tracking configurations using CKF or mkFit for track fitting. Each measurement is carried out on a dedicated machine equipped with two AMD EPYC ``Milan'' 7763 CPUs, for a total of 256 cores (2 sockets $\times$ 64 physical cores $\times$ 2 logical cores per core), and two NVIDIA T4 GPUs. Results are obtained by repeating the timing measurement three times and taking the average. To emulate the 50-50 split of processing power between CPU and GPU targeted for Phase-2 HLT, 8 parallel jobs, each with 16 threads and 16 streams, are restricted to run on a single CPU socket. When the configuration involves GPU processing, both GPUs are used. Before each measurement, input files are read into memory and cached by the operating system to emulate data-taking conditions. The timing is evaluated on 1000 simulated $t\bar{t}$ events with an average pileup of 200.\\

When GPU accelerators are used, the total tracking time is of approximately 700 ms per event; out of these, roughly 300 ms are taken by the legacy CKF track fitting~\cite{CMS-DP-2026-026}. When mkFit is used for track fitting, the combined track building and fitting processing time is reduced by approximately 220 ms, corresponding to a 71\% reduction in the track fitting time, including data-format conversions. This translates into an overall tracking time reduction of 32\% (14\%) in the realistic scenario where GPU accelerators are (not) used. The processing time for track building, track fitting and conversion steps of the HLT tracking sequence, that run only on CPU, are summarized in Table~\ref{tab:timing_breakdown}, while the time reduction achieved thanks the the usage of mkFit for track fitting in different scenarios is summarized in Table~\ref{tab:timing_summary}.

\begin{table}[h!]
\centering
\caption{Processing time for the track building, track fitting, and conversion steps of the HLT tracking sequence, for the CKF and mkFit final fit configurations, measured on 1000 simulated $t\bar{t}$ events with an average pileup of 200.}
\label{tab:timing_breakdown}
\begin{tabular}{lcccc}
\hline
 & Track & Conversion & Track & Conversion \\
 & building & (candidate) & fitting & (track) \\
\hline
CKF final fit   & 92 ms & 18 ms & 300 ms & --    \\
MkFit final fit & 92 ms    & --    & 50 ms  & 42 ms \\
\hline
\end{tabular}
\end{table}

\begin{table}[h!]
\centering
\caption{Summary of the fitting and conversion time, and the total tracking time with and without GPUs for the CKF and mkFit final fit configurations, together with the relative reduction achieved thanks to the usage of mkFit for track fitting.}
\label{tab:timing_summary}
\begin{tabular}{lccc}
\hline
 & Fitting + & Full tracking & Full tracking \\
 & Conversion & (with GPUs) & (CPU only) \\
\hline
CKF final fit   & 318 ms & 708 ms & 1687 ms \\
MkFit final fit & 92 ms  & 484 ms & 1456 ms \\
\hline
Reduction       & 71\%   & 32\%   & 14\%   \\
\hline
\end{tabular}
\end{table}

\section{Conclusions}
\label{conclusions}

The adoption of mkFit for track fitting is motivated both by its demonstrated computational performance during the Run 3 of the LHC and by the need to further reduce tracking time ahead of the HL-LHC data taking. Good physics performance and timing improvements have been shown in a realistic Phase-2 simulation of the CMS HLT reconstruction, using the new tracking configuration proposed for CMS, with the mkFit final fit used in place of the legacy CKF algorithm for comparison. The timing performance improves significantly the HLT tracking, yielding a 32\% reduction of the overall tracking time in a realistic scenario where GPU accelerators are used. Further details are to be finalized in the upcoming future for a complete assessment of the performance, with the goal to use mkFit for track fitting during Run 4.\\

\small{This work was supported by the National Science Foundation under Cooperative Agreements OAC-1836650 and PHY-2323298.}

\end{document}